\documentclass[english,11pt]{article}

\usepackage{changes}
\usepackage{babel}
\usepackage[T1]{fontenc}
\usepackage[latin9]{inputenc}
\usepackage{geometry}
\usepackage{dcolumn} 
\newcolumntype{d}[1]{D{.}{.}{#1}}
\usepackage{color}
\usepackage{array,multirow,tablefootnote,float,floatpag,booktabs,makecell,longtable}
\usepackage{placeins}
\usepackage{nicefrac}
\usepackage{caption}
\usepackage[labelformat=simple]{subcaption}
\usepackage[flushleft]{threeparttable}
\usepackage{amstext}
\usepackage{amsthm}
\usepackage{graphicx}
\usepackage{enumitem}
\usepackage[authoryear]{natbib}

\ifx11 
    \newcommand{\bisinm}[1]{\todo[color=blue!30,caption={}]{\textbf{MV:} #1}}
    \newcommand{\bisini}[1]{\todo[inline,color=blue!30,caption={}]{\textbf{MV:} #1}}
    \newcommand{\andream}[1]{\todo[color=green!30,caption={}]{\textbf{AM:} #1}}
    \newcommand{\andreai}[1]{\todo[inline,color=green!30,caption={}]{\textbf{AM:} #1}}
\else
    \newcommand{\bisinm}[1]{}
    \newcommand{\bisini}[1]{}
    \newcommand{\andream}[1]{}
    \newcommand{\andreai}[1]{}
\fi

\usepackage[unicode=true,pdfusetitle,
 bookmarks=true,bookmarksnumbered=true,bookmarksopen=true,bookmarksopenlevel=2,
 breaklinks=true,pdfborder={0 0 0},pdfborderstyle={},backref=page,colorlinks=true]
 {hyperref}
\definecolor{darkblue}{rgb}{0,0.1,0.5}
\hypersetup{citecolor=darkblue,linkcolor=darkblue,urlcolor=darkblue}

\renewcommand*{\backref}[1]{}
\renewcommand*{\backrefalt}[4]{{\ifcase #1%
          \or(Cited on page~#2)%
          \else(Cited on pages #2)%
    \fi%
    }}
    
\newcommand{\lineLd}{The horizontal lines indicate the proportion of infected triggering a lockdown}
\newcommand{\lineLdReop}{\lineLd\ or a reopening}
\newcommand{\tablenote}{The figure illustrates the dynamics of active cases ($A+Y$) as a fraction of the total population under different lockdown policies. }
\newcommand{\tablenoteLC}{The figure illustrates the dynamics of active cases ($A+Y$) as a fraction of the total population with naive policymaking. }
\newcommand{\tab}{The table reports steady-state outcomes as a fraction of total population and the number of infected by location as a fraction of the total number of infected $D+R$}

\begin{document}
\setcounter{footnote}{2}
\author{
    Alberto Bisin
    \and 
    Andrea Moro}

\title{Spatial-SIR with Network Structure and Behavior: \\ Lockdown Rules and the Lucas Critique\thanks
{
    Please check our websites for an updated version of this paper.  
    Bisin:  New York University, \texttt{\href{https://wp.nyu.edu/albertobisin/}    {wp.nyu.edu/albertobisin/},               
        \href{mailto:alberto.bisin.nyu.edu}{alberto.bisin@nyu.edu}%
    }. 
    Moro:  Vanderbilt University 
    \texttt{
        \href{https://andreamoro.net}{andreamoro.net}, 
        \href{mailto:andrea@andreamoro.net}{andrea@andreamoro.net}.
    }
}
\date{\today \\ {\small First version: August 4, 2020}}
}
\maketitle
\begin{abstract}
    We introduce a model of the diffusion of an epidemic with demographically heterogeneous agents interacting socially on a spatially structured network.
    Contagion-risk averse agents respond behaviorally to the diffusion of the infections by limiting their social interactions. Schools and workplaces also respond by allowing students and employees to attend and work remotely. The spatial structure induces local herd immunities along socio-demographic dimensions, which significantly affect the dynamics of infections.  
    We study several non-pharmaceutical interventions; e.g., i) lockdown rules, which set thresholds on the spread of the infection for the closing and reopening of economic activities; ii) neighborhood lockdowns, leveraging granular (neighborhood-level)  information to improve the effectiveness public health policies;  iii) selective lockdowns, which restrict social interactions by location (in the network) and by the demographic characteristics of the agents.
    Substantiating a ``Lucas critique'' argument, we assess the cost of naive discretionary policies ignoring agents and firms' behavioral responses.
\end{abstract}

\newpage
\section{Introduction}

In this paper, we develop a  
diffusion model of infection to study the effectiveness of different public health policies
on contagion dynamics. We extend the Spatial-SIR model introduced in \cite{bisinmoro2020Geo}  to account  for a demographic and a network structure of the social contacts  driving the diffusion process. This allows us to interact population and location heterogeneity with the spatial dimension of contacts. We also account for agents' and firms' behavioral responses to the diffusion of the epidemics and to policy interventions. 

In the model,
agents belong to three demographic types: Young (students/workers),  Not employed, Old. Each day, the types potentially visit different locations in a structured network: the City, School/Work locations, and Home, which differ by contact density. Consequently, the model displays heterogeneous rates of diffusion of the epidemic by demographic type and across the network of locations. Furthermore, depending on the spread of the epidemic, contagion-risk averse agents may reduce their social interactions, and firms may instruct their workers to operate remotely. We calibrate the model to reproduce several stylized facts about the diffusion of SARS-CoV-2 epidemics. We calibrate agent types and family compositions to match the U.S. demographics, particularly the family size distribution by age. Next, we simulate the calibrated model.  

The model simulation identifies several important factors relevant to understanding
the heterogeneous rates of diffusion of the epidemic, highlighting that local herd immunities may form not
only geographically (the focus in \cite{bisinmoro2020Geo}) but also along socio-demographic dimensions depending on the network structure.
For instance, the City originates more infections than School/Work and School/Work more than Home, even though Home is the densest location and the City is the least dense. This result occurs because the ranking of the locations in terms of density is opposite to the ranking in terms of size, therefore herd immunity is achieved earlier at Home and School/Work than in the City. Similarly, the epidemic is primarily concentrated on the Young because they are exposed at School/Work, where there are more contacts than at Home. These effects, which we say arise from \emph{local herd immunity} are akin to a selection effect, whereby the epidemic spreads selectively by location and demographic structure. 

Our analysis also illustrates the relevance of the indirect effects induced by agents' and firms' behavioral responses across locations and demographic types. For instance, the Youngs' behavioral responses, preventing exposure in School/Work locations, have a sizable substantial effect on the Old in the City, where most of the Old get infected. 

We use this model to study the effects of several  
public health policy
interventions. Simulations allow us to evaluate the outcomes of these interventions with respect to, e.g.,  the peaks of the first and second waves of infection and the spread of the infection by demographic structure. Specifically, we emphasize public health policy rules, that is, articulated interventions on an epidemic's whole course possibly developing over various waves. We study, e.g., lockdown rules, where the public health authorities set thresholds levels of active infection cases determining when social interactions are restricted and when they are allowed. Lockdown rules interact in rather subtle ways with the dynamics of herd immunity. When the lockdown is placed too early with respect to the spread of the epidemic, the fraction of infected is too small, herd immunity is too far out, and a second wave of infections at reopening reaches a peak higher than the first one. If reopening is delayed to occur at a lower fraction of active cases than the one that triggered the lockdown,
a strategy we labeled \emph{cautious reopening},
herd behavior is more advanced upon reopening, substantially dampening the second wave. 

We also simulate the ability of the government to exploit granular information on the epidemics by imposing lockdowns at the neighborhood level based on the local progression of the infection. Neighborhood-level lockdown policies can flatten the curve of infections  and generate fewer infected at steady state, if compared to a city-wide lockdown. However, the effectiveness of these localized lockdowns crucially depends on how segregated neighborhoods are, in particular on the level of interaction across neighborhoods caused by agents visiting School/Work locations.

Moreover, we find that when neighborhoods are heterogeneous in their demographic structure, the location of the initial cluster is a crucial determinant of the infection dynamics and steady-state outcome. When the infection starts in a neighborhoods with smaller average household size, the infection progresses more slowly; therefore, it has time to spread to nearby neighborhoods, generating a higher city-wide peak and steady-state levels of infections. When the infection starts in neighborhoods with large families, it affects locally a large fraction of people, achieving quickly a local herd immunity that spares, to some extent, the rest of the city from a higher level of infections.

The model's network and demographic structure also allow us to study various selective lockdown policies, where lockdowns are imposed by the location of social interactions and/or by the agents' demographic characteristics. We concentrate on two such policies.  In the first one, we limit the lockdown to the City, and we let firms/schools decide whether to operate remotely. In the second one, we limit the lockdown to the Old, a policy aiming at reducing total fatalities while bearing limited economic costs (the Old are not economically active but suffer a higher fatality rate than other demographic groups). 
Both these policies have the important property that they are much less costly economically because workers/students remain active. This benefit is traded-off with health-outcome costs at the steady-state or at the peak (relevant when there is a health care resource constraint, such as hospital/ICU beds).  While these selective lockdowns have positive direct effects in terms of economic costs, indirect effects across locations and/or demographic types could, in principle, substantially limit their advantage over general lockdowns. This outcome does not appear to occur in our simulations. The City-Only lockdown does not induce a much larger fraction of infected at steady-state than the general lockdown, while the Old-Only lockdown is very successful in limiting Old agents' deaths. 

Last but not least, 
we illustrate the implications of what economists refer to as the Lucas Critique in the context of epidemiological models. 
Policy evaluations and interventions disregarding that agents' and firms' 
``decision rules vary systematically with changes in the structure of series relevant to the decision-maker'' (\cite{lucas1976econometric}) 
might lead to policy decisions that are very costly in terms of their effects on the epidemic dynamics.
The costs of disregarding behavioral responses are clearer if we consider that the lockdown and opening thresholds are calibrated to flatten the infection curve to avoid hitting a constraint on the available health care resources. 
In this case, our simulations imply under-utilization of health care resources after lockdown (that is, the lockdown will turn out to be stricter than necessary) and possibly an over-run of these resources after reopening. 

\section{Epidemic by Spatial, Demographic, Network, and Behavior Characteristics} \label{sec:model}

We introduce a model of the dynamics of an epidemic which, while stylized, contains several important components needed to analyze the effects of different forms of lockdowns.  Our goal is to maintain a parsimonious structure while replicating 
the features of general epidemics that are relevant for policy analysis, to produce relatively robust qualitative results.\footnote{Frontier research in epidemiology has proposed  models of contagion, extending the classic SIR model in many directions, including detailed descriptions of the geographic or network characteristics of the agents' social interactions - see, e.g., \cite{eubank2004modelling} and 
 the research at  \href{https://covid19.gleamproject.org}{GLEAM project}, 
\href{https://www.mobs-lab.org/projects.html}{mobs-lab}, and the \href{https://www.imperial.ac.uk/mrc-global-infectious-disease-analysis}{Imperial's college MRC Centre for Global Infectious Disease Analysis}. While these are large-scale models fundamentally aimed at forecasting with accuracy and precision, their complexity makes it difficult to uncover the main driving forces of the epidemics and policy interventions' effects. } 

We build on the  Spatial-SIR model we have introduced in \cite{bisinmoro2020Geo}. In the Spatial-SIR, agents are located and move in a two-dimensional space. With some probability, they become infected when they get close to an infected agent.\footnote{Several recent contributions to the study of the SARS-CoV-2 epidemic in economics have restricted their epidemiology component to the SIR model, without modeling the spatial dimension of the epidemics;  see e.g.,   \cite{fernandez2020estimating},    \cite{atkeson2020will}, \cite{eichenbaum2020macroeconomics}, \cite{giannitsarou2020waning}, \cite{keppo2020behavioral}, \cite{rowthorn2012optimal} \cite{weitz2020moving}.  
As exceptions, \cite{antras2020globalization}, \cite{brotherhood2020slums}, 
\cite{cua2020structural}, and   \cite{glaeser2020much} introduce connections between different geographical units. \label{geo} }     In this paper, we introduce heterogeneity in agents' type (by age and occupation status) and in the characteristics of the social spaces where they interact so that they interact in distinct networks.\footnote{Along these lines,  \cite{acemoglu2020multi}, \cite{alfaro2020social}, \cite{sel-Babajanyan}, \cite{sel-baqaee}, \cite{sel-baqaee2020policies}, \cite{bognanni2020economic},
\cite{brotherhood2020economic}, \cite{sel-fischer}, \cite{sel-giagheddu2020}, \cite{sel-gollier2020cost} add a demographic structure like we do;  \cite{Azzimontietal2020} adds a network structure;  \cite{ellison}
  allow for heterogeneity of the contact process between subpopulations.}

We model agents responding to the dynamics of the epidemic, by choosing to limit their social contacts, and we assume that some schools/firms adopt behavioral responses similar to the agents', that is, they choose to shut down operations or to allow workers to work remotely, to limit infections in the School/Work locations.\footnote{Models of rational agents limiting contacts to reduce the risk of  infection  include \cite{fenichel2013economic} \cite{weitz2020moving} in epidemiology (see also \cite{funk2010modelling} and  \cite{verelst2016behavioural}  for surveys); and  
\cite{geoffard1996rational}, \cite{goenka2012infectious}, 
\cite{acemoglu2020multi}, 
\cite{aguirre}, \cite{argente2020cost}, 
\cite{bethune2020covid}, \cite{farboodi2020}, \cite{fernandez2020estimating}, 
\cite{greenwood2019equilibrium},
\cite{keppo2020behavioral},   \cite{toxvaerd2020equilibrium} (as well as several of the papers cited in Footnote \ref{geo}, modeling spatial extensions of SIR) in economics. Empirically, several papers use cellular phone data to document how fear of contagion has affected mobility or show that the economy deteriorated substantially even before or independently of lockdown orders. See, for example,
\cite{alfaro2020social},
\cite{aum2020covid}, 
\cite{barrios},
\cite{bartik2020small}, 
\cite{cicala},
\cite{coibion2020cost}, 
\cite{goolsbee2020fear},
\cite{gupta2020effects},
\cite{NBERw27061}, 
\cite{rojas2020cure}.}  

\subsection{State transitions}
Let $\mathcal{S}=\{S, A, Y, R, D\}$ denote the individual state-space, indicating Susceptibles, Asymptomatic, sYmptomatic, Recovered, and Dead.\footnote{There is no accepted denomination for the form of compartmental model we adopt in this paper. Alternative denominations would include A-SIRD and Stochastic Symptom-Response SIRD, following, respectively, \cite{gaeta2020simple, leung2018infector}.} We assume  the following transitions: 
i) a Susceptible agent in a location within distance $p$ from the location of an Asymptomatic becomes infected with
probability $\pi$;\footnote{We assume sYmptomatic agents are isolated at home or in the hospital, therefore not contagious.}
ii) an Asymptomatic agent infected
at $t$, at any future period, can become 
sYmptomatic with probability $\nu$, or can Recover with probability $\rho$; 
iii) a sYmptomatic agent can Recover or Die at any period with probabilities $\rho$ and $\delta$, respectively;  
iv) Dead and Recovered agents never leave these states.  

\subsection{Demographic and network structure}
There are three \emph{types of} locations: the \emph{City} at large, a set of \emph{School/Work} locations shared by subsets of agents, 
and  \emph{Home} locations shared by family members.   
Agents differ by age and occupation in their access to these three types of locations.  

There are three {\em demographic types}  of agents:\footnote{The age structure of the population has specific relevance for the SARS-CoV-2 epidemic because the case-fatality rate is very skewed, higher for older agents.}  the \emph{Young}, the \emph{Not employed,} and the \emph{Old}. The Young represent students/workers who spend part of their day in School/Work, that is, in a location where they interact daily with the same individuals. The Not employed represent adult individuals below 65 years of age that do not have access to a School/Work location. The Old are not working but differ from the Not employed because of their higher fatality rate and because they tend to live at Home with individuals of the same type. 
The demographic structure interacts with the location structure in that the Old do not visit School/Work locations. 

\subsection{Social interactions}

Each day is divided into three periods dedicated to the location where agents interact: Period 1 is devoted to visiting the City, Period 2 to visiting the School/Work location and Period 3 to stay at Home. However, agents might choose or be structurally restricted from visiting the City and/or School/Work and hence stay Home, depending on their demographic and/or health state.\footnote{Agents might be restricted to visit some location also as a policy choice. We discuss these restrictions in the next section.}
SYmptomatic agents must stay at Home in all three periods. If not sYmptomatic, all agents visit the City in Period 1, and all Young agents visit the School/Work location in Period 2, when the other agents stay Home. Table \ref{location} summarizes the locations visited by agents by the time of day.

\begin{table}[t]
   \centering
    \caption{Location by time of day}\label{location}
   \renewcommand{\arraystretch}{1.2} 
    \begin{tabular}{l>{\centering}m{0.2\linewidth}>{\centering}m{0.2\linewidth}>{\centering\arraybackslash}m{0.2\linewidth}}    
    \toprule
        Location type & Period 1 & Period  2  & Period 3 \tabularnewline
    \midrule
        City & All except sYmptomatics  & No-one &  No-one  \tabularnewline 
        School/Work & No-one & Young except sYmptomatics & No-one \tabularnewline 
        Home & sYmptomatics  & sYmptomatics, Not employed, Old & All \tabularnewline
    \bottomrule
    \end{tabular}
\end{table}

Agents move randomly in the City: Every day $t=[0,T]$, agents travel distance $\mu$ 
toward a random direction of $d\sim U[0,2\pi]$ radians. By doing so, they potentially meet new individuals every day; therefore, $\mu$ models the speed at which agents find new contacts, potentially in a different state than their previous neighbors.
Each Young is randomly assigned to a School/Work location and does not move within it. School/Work locations are far from each other (there is no possibility of contagion across School/Work locations). Finally, agents are assigned to families located in Homes that are far from each other and far from School/Work locations. 

The distinction we introduce across the three locations abstractly captures different relative densities in agents' social interactions. The School/Work location represents a non-random location that is visited every day by the same set of people and is denser than the City (see the next section for details on how location density is calibrated). Home represents the location with the tightest interactions: agents from the same family have identical Home locations; that is, they are necessarily in contact with each other.  On the other hand, at School/Work and in the City, the contagion may occur only if people are close in space. 
 \begin{table}[t]
 	\thisfloatpagestyle{empty}
	\caption{Calibrated parameter values}\label{parameters}  
    \centering
    \renewcommand{\arraystretch}{1.2} 
    \begin{tabular}{ll}
        \toprule
        Parameter                       & Value   \tabularnewline \midrule 
        Number or people                    & 26,600        
        \\ Initially infected                & 30
        \\
        \multicolumn{1}{l}{Family size and age distr.}  & \multicolumn{1}{l}{match 2018 ACS} 
        \tabularnewline
        School/Work population (avg.)  & 100 
        \\ Group quarter population              & 50  
        \\ School/Work side length             & 0.003304
        \tabularnewline                                            
        Contagion radius                & 0.00805  
        \\ Distance traveled in City         & 0.028175      
        \tabularnewline 
        Prob. of becoming symptomatic   & 0.09  
        \\   Prob. of recovery            & 0.05   
        \tabularnewline
        Fatality rate Young & 0.000533 
        \\ Fatality rate, Old      & 0.00533
        \tabularnewline    
        Contagion probability, Home & 0.064 \\ 
        Contagion probability, City School/Work & 0.032 
        \tabularnewline
        Behavioral responses, agents & $\phi:  .88, \underline{I}:.001$ 
        \\ Behavioral response, firms & $\phi:.88, \underline{I}: .00062 $
        \tabularnewline
        \bottomrule
    \end{tabular}

	\medskip
	\caption*{\normalfont \footnotesize  
	\emph{Note:} ACS: American Community Survey. Group quarters are places where people live or stay in group living arrangements. The School/Work side length is relative to the City's side length, which is normalized to 1. See Appendix \ref{app:calibration} for details.
	}

\end{table}

\subsection{Behavior}\label{sec:behavioral}


In this section, we model agents responding to the epidemic dynamics by choosing to limit their contacts.  
Following \cite{keppo2020behavioral}, we introduce a reduced-form behavioral response,  represented by a function $0 \leq \alpha( I_t) \leq 1$, representing the fraction of agents who choose to interact in the City as a function of the number of infected $I_t$:\footnote{In our simulations, we assume the behavioral response depends on the fraction of Asymptomatics because the sYmptomatics are isolated at Home. Since the fraction of Asymptomatics is not observable, the behavioral response could only depend on the number of sYmptomatics, as a proxy; with Rational Expectations, however,  the agents know (rationally infer) the equilibrium map from $Y$ to $A$, say $A(Y)$, possibly with noise; see \cite{BisinMoroRat2020} or \cite{gans2020}.}

\begin{equation} 
    \alpha(I_{t})=\left\{ 
    \begin{array}{ll}   
        1  & if \; \;  I_{t} \leq \underline{I} \\ 
        \left( \frac{\underline{I}}{I_{t}} \right)^{1-\phi}  & if \; \;  I_{t}>\underline{I}
    \end{array} \right. .  
    \label{Ketal}
\end{equation}

We rank the agents (and School/Work locations) by contagion-risk aversion. Agents choosing not to interact in the City (and School/Firms leaving agents at Home) are the most contagion-risk averse. 
We also assume some firms and schools adopt behavioral responses similar to the agents' according to (\ref{Ketal}), that is, they choose to shut down operations or to allow workers to
work remotely, to limit infections at school or work. 

\section{Calibration and simulation }

We  use data from the 2018 American Community Survey (ACS) to classify agents by type and allocate them to households replicating the size and age distribution of 
American households. We calibrate the three locations' relative density to match the data on the distribution of contacts reported by \citet{Mossong_2008}. Transition probabilities between states match various SARS-CoV-2 parameters from epidemiological studies, notably, e.g., \cite{ferguson2020report}. We calibrate contagion rates and the agents' movement speed in the City to fit the epidemic's initial growth rate, proxied by the growth rates of deaths in Lombardy. Finally, behavioral response parameters are calibrated following \cite{keppo2020behavioral}.\footnote{We acknowledge the uncertainty in the literature concerning many epidemiological parameters pertaining to this epidemic. Our approach is less damaging when aiming at understanding mechanisms and orders-of-magnitude rather than at precise forecasts.} Appendix \ref{app:calibration} describes the calibration in detail. 
Table \ref{parameters} reports the parameter used in our main specification.   

\begin{figure}[t]
    \caption{Simulation of the benchmark model\label{exh:behavioral}}
    \makebox[\textwidth][c]{ 
    \begin{subfigure}{0.35\textwidth}
        \centering
       \includegraphics[width=\textwidth]{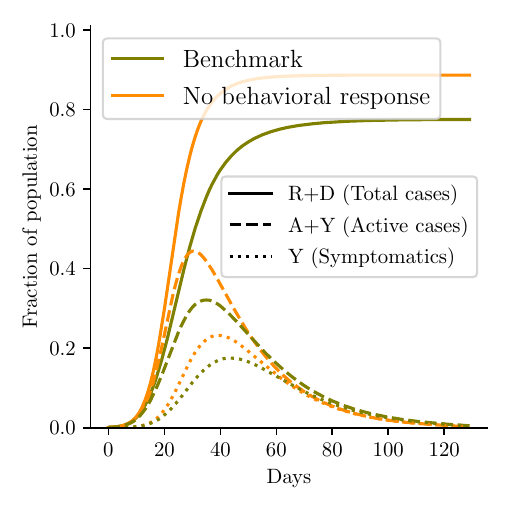}
       \caption{Infection dynamics}\label{fig:behavioral}
    \end{subfigure}%
    \begin{subfigure}{0.35\textwidth}
        \centering
       \includegraphics[width=\textwidth]{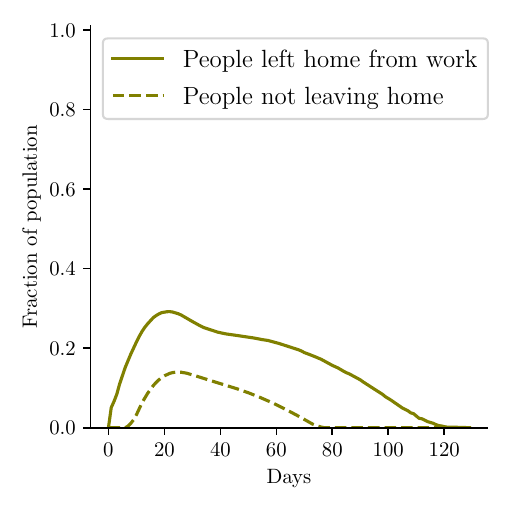}
       \caption{Behavioral responses}\label{fig:respsize}
    \end{subfigure}%
    \begin{subfigure}{0.35\textwidth}
        \centering
       \includegraphics[width=\textwidth]{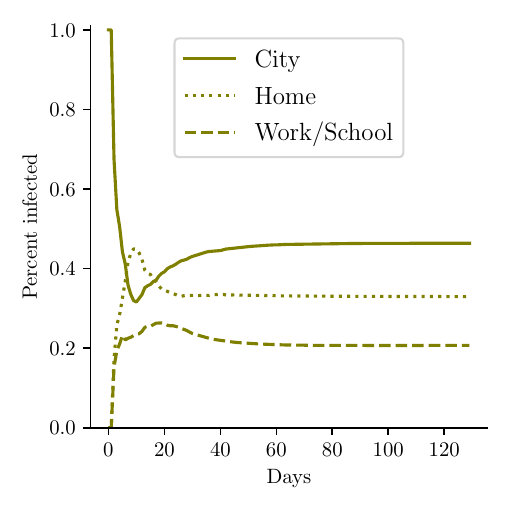}
       \caption{Share of infections}\label{fig:infloc}
    \end{subfigure}
    }
    
    \caption*{\normalfont \footnotesize 
        \emph{Note:} Infection dynamics, Benchmark model (green lines), and model without behavioral responses (orange lines)}
\end{figure}
Results from the simulation of the benchmark model are illustrated 
in Figure \ref{exh:behavioral};  Figure \ref{fig:behavioral} reports the epidemic dynamics; Figure \ref{fig:respsize} describes the behavioral response, reporting the fraction of agents (out of the total population) choosing not to visit the city (dashed line) and the fraction of agents left home from work because their workplace shuts down; and Figure \ref{fig:infloc} the locations of the infections. Table \ref{tab:behavioral} reports the steady-state outcomes.\footnote{\label{20randomreps}All results are obtained averaging over 20 replications of the simulation, randomizing over the initial location of individuals in the City and at School/Work}, including the locations of infection by agent type and the percent of the population in three different states at the steady-state.

Generally, our simulation of the model demonstrates that {\em local herd immunity} - which in \cite{bisinmoro2020Geo} we argue explains the dependence of the dynamics of the epidemic from geographical characteristics - also explains the dependence of the epidemic from location and demographic characteristics. With respect to location, the City originates more infections, 46 percent (vs. 21 percent in School/Work location, and 33 percent at Home) even though Homes and School/Work locations are denser because more contacts with different people are generated in the City. Herd immunity is achieved  {\em locally} earlier at Home and at School/Work than in the City. The infection dynamics by location shows a larger fraction of infections at Home than at School/Work and in the City after about ten days, when the City relative infection rate starts to pick up.\footnote{We set the location of infection at $t=0$ in the City.} 

More Young are infected because of their exposure to more contacts at School/Work than at Home. Old and the Not employed are hit differently because of family structure: the latter live in households with relatively more young individuals that are infected at School/Work locations. These effects are akin to a selection effect, whereby the epidemic spreads selectively by location and demographic structure.\footnote{See \cite{gomes2020individual} and \cite{britton2020mathematical} for  related theoretical analyses.}

\begin{table}[t]
    \caption{Benchmark model, infection location and steady state outcomes}
    \label{tab:bench}
    \begin{subtable}{\linewidth}
 	\centering
 	\bigskip
 	
	\begin{tabular}{m{0.20\linewidth}
	 *5{>{\centering}m{0.07\linewidth}}
	 >{\centering \arraybackslash}m{0.07\linewidth}}
        \toprule
 & \multicolumn{3}{c}{Steady-state outcomes} & \multicolumn{3}{c}{Share of infections}\tabularnewline
& D & R & Peak A+Y & City & W/S & Home  \tabularnewline
        \midrule
        All     & 0.008         & 0.766         & 0.323         & 0.463         & 0.205         & 0.332 \tabularnewline
        \midrule
        Young          & 0.004         & 0.819         & 0.349         & 0.394         & 0.296         & 0.309 \tabularnewline
        Not employed           & 0.004         & 0.728         & 0.290         & 0.537         & -         & 0.463 \tabularnewline
        Old            & 0.029         & 0.599         & 0.249         & 0.726         & -         & 0.274 \tabularnewline
         \midrule
 	\end{tabular}
 	\caption{Benchmark model
		\label{tab:behavioral}} 
    \end{subtable}
    \bigskip
    
    \begin{subtable}{\linewidth}
        \centering
	\begin{tabular}{m{0.20\linewidth}
	 *5{>{\centering}m{0.07\linewidth}}
	 >{\centering \arraybackslash}m{0.07\linewidth}}
            All     & 0.010         & 0.876         & 0.446         & 0.472         & 0.294         & 0.234 \tabularnewline
    \midrule
             Young          & 0.005         & 0.933         & 0.486         & 0.372         & 0.427         & 0.201 \tabularnewline
             Not employed           & 0.005         & 0.815         & 0.383         & 0.626         & -         & 0.374 \tabularnewline
             Old            & 0.036         & 0.719         & 0.365         & 0.777         & -         & 0.223 \tabularnewline
     \midrule
    	 \end{tabular} 
	 \caption{Benchmark model without behavioral responses}\label{tab:infloc}
    \end{subtable}
    \caption*{\normalfont \footnotesize \emph{Note:} \tab.}
\end{table}

It is interesting to compare the simulation of the benchmark model to a model simulated without behavioral responses (see the orange lines in Figure \ref{fig:behavioral} and Table \ref{tab:infloc}). The total fraction of Recovered and Dead at the Steady-state is 0.77 in the benchmark model but would have been .88 without the behavioral response. The fraction
of infections occurring at Home is almost ten percentage points lower without behavioral response. The symmetric opposite effect is mostly concentrated at School/Work rather than in the City, whose fractions of infection increase, respectively, by 9 and 1 percentage points.   
The total rise in infections at School/Work affects mostly the Young, who get infected less at Home (from 31 to 20 percent)  and more at School/Work (from 30 to 43 percent). 
At the steady-state, the proportion of Young Recovered or Dead increases from 82 percent to 94 percent. 

The most important implication resulting from this simulation is the indirect effect of the behavioral response across locations and demographic types. Specifically, the behavioral responses of (mostly) the Young, (mostly) in the School/Work location have a sizable important effect on the Old in the City  (the Old get infected mostly in the City, over 70 percent, with or without behavioral response). Notably, the compensating effect of higher infections at Home with the behavioral response is of reduced importance for the Old because of family composition (the Old tend to live with other Old). At the steady-state, the proportion of Old Recovered or Dead increases from 63 percent to 73 percent.\footnote{A simulation - which we do not report in the text - where we only allow for a behavioral response by firms, has a much smaller effect on the Old, suggesting that the main impact on the Old goes through the reduction of infections in the City.} The impact of behavioral response on the Not employed is instead moderate because their family composition induces them to be infected more at Home. 

Finally, we assess the relevance of the spatial dimension, which induces local herd immunity,  in conjunction with the network structure, which distinguishes City, School/Work and Home. 
To this end, we compare our benchmark spatial model with one characterized by random matching of agents, that is, with no spatial structure, but with the network structure. In the random matching model, each agent can be infected by anyone in the City and in the School/Work every day.\footnote{Specifically, we expand the contagion radius so that the contagion circle includes the entire City. Similarly, we expand the contagion ratio in each  School/Work to include the entire establishment.  We recalibrate the contagion probability in the City and in each School/Work so that the average number of people one infected agent infects is kept constant across the benchmark spatial and random matching models.}  

\newcommand{\mc}[1]{\multicolumn{1}{c}{#1}}
\begin{figure}[!t]
    \centering
    \caption{Benchmark model comparison with random matching    \label{fig:random}}
    \begin{subfigure}{0.35\textwidth}
        \centering
       \includegraphics[width=\textwidth]{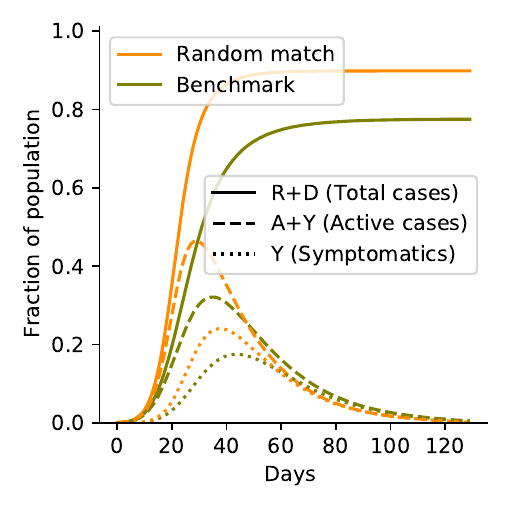}
       \caption{Infection dynamics}\label{fig:r-dyn}
    \end{subfigure}%
    \begin{subfigure}{0.35\textwidth}
        \centering
       \includegraphics[width=\textwidth]{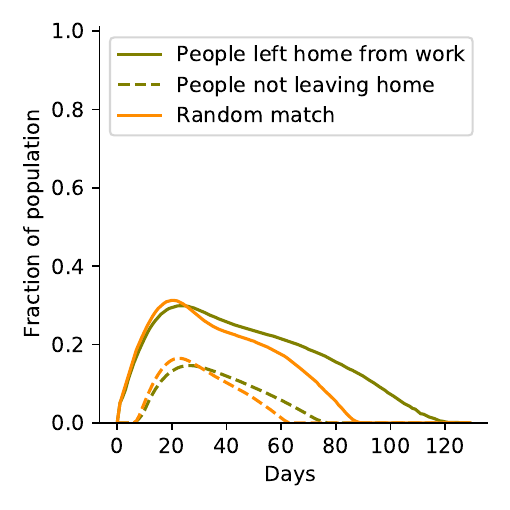}
       \caption{Behavioral responses}\label{fig:r-beh}
    \end{subfigure}%
    \medskip 

    \begin{subfigure}{\linewidth} \centering
    \begin{tabular}{m{0.16\linewidth}
                    *3{>{\centering}m{0.16\linewidth}}
                    >{\centering \arraybackslash}m{0.16\linewidth}}
    	\midrule 
    	 & \multicolumn{4}{c}{Fraction infected (R+D) in steady-state}\tabularnewline

  & \mc{City} & \mc{Work/School} & \mc{Home} & \mc{Total} \tabularnewline 
       \midrule 
All               & 0.355 (--.004)  & 0.152 (--.007) & 0.390  (+.133) & 0.898 (+.124)  \tabularnewline
\midrule
 Young           & 0.283 (--.042)  & 0.234 (--.010) & 0.440 (+.185) & 0.957 (+.133) \tabularnewline
 Not employed    & 0.459 (+.066)  & -     & 0.359 (+.020)   & 0.818 (+.086) \tabularnewline
 Old             & 0.523 (+.067)  & -     & 0.227 (+.055)  & 0.750 (+.122) \tabularnewline
 	    \midrule
 	\end{tabular} \smallskip
 	
 	\caption{Fraction infected by location in random matching model \\ and difference with benchmark model}\label{tab:random}
 	    \end{subfigure}
 	\medskip
    \caption*{\normalfont \footnotesize 
        \emph{Note:} The figures illustrate the infection dynamics, Benchmark model (green lines), and random-mathing model  (orange lines). The table reports the number of infected in the random matching model as a fraction of the total population of each type, by location of infection. Difference between random matching model and benchmark model in parenthesis.}
\end{figure}
 	   
Results are reported in Figure \ref{fig:random}. 
The spatial component of our analysis matters first of all in that the steady state of  infections  is greater with random matching, as shown in Figure \ref{fig:r-dyn}. Importantly, this effect is not  crucially dependent on behavior, as it is clear from Figure \ref{fig:r-beh}
that the behavioral reactions of both firms (leaving workers home) and agents (staying home from the City) is only slightly more pronounced under random matching. It is rather the local herd immunity phenomenon in our benchmark model which is fundamentally responsible for the reduction of the steady state fraction of infected with respect to the random matching model. The mechanism operates  however through a rather subtle implication of the interaction of the spatial component of our benchmark model with the network structure. This is  apparent from the distribution of infections by type of agent and by  location reported in the table underneath Figure \ref{fig:random} (panel \ref{tab:random}).  Indeed, even though random matching operates only in the City, and all types interact in the City, the extra infected are largely at Home (+.133 percentage points) and are mostly the Young (+.133). Furthermore, the effect is larger for the Old (+.122) than for the Not employed (+.086), even though these two types are exposed in the same way to interactions in the City.\footnote{We thank one referee for pointing out this piece of evidence.}  
In fact, the first order effect of the lack of  local herd immunity in the City with random matching  is to speed up the infection process in the early stages of the epidemic. This faster spreading of  infections in the City in turn speeds up the infections at Home, at an even accelerating rate due to the fact that infections at Home are much faster than at any other location since agents live in close quarters (at zero distance from each other in the simulations). 
Of course, agents who are infected at Home are not infected in other locations, in particular not in the City. This explains why the effects of random matching are seen mostly at Home rather than in the City. 
This effect is more prominent for the Young because the mechanism we are illustrating is accelerated in their case by the fact that they can also be  infected in Work/School. 
Furthermore, these effects are larger for the Old than for the Not employed because the demographic composition of the Home is such that the Old live mostly with other Old. In the benchmark model, the effect of local herd immunity is tamed for the Not employed because they are more likely to live with the Young, who infect them at Home even when there are no active cases locally. When local herd immunity is removed by random matching, the infections at Home increase more for the Old than for the Not employed.

\section{Lockdown policies}\label{sec:lockdowns}

This section explores the effects of various non-pharmaceutical interventions along the lines of those proposed and implemented by various administrations during the COVID-19 outbreak. Our
goal is to exploit the spatial nature of the model, the structure of the network of interactions, and the (distinction by) demographic characteristics of the agents to analyze the different impact of these policies both in the aggregate and across types. 
 
We study general lockdown rules, that is, interventions where the lockdown and reopening rules depend on the spread of the infections. We also study selective lockdowns, where only the City is locked down, or only Old individuals are restricted in movement across locations (either confined either at Home or in quarantine facilities of different sizes). In all the cases we study, policies overlap with the behavioral response; that is, they are applied
to the population of agents and firms randomly, not accounting for their behavioral response. In other word, lockdown policies  may apply 
to agents (resp. firms) that 
would have chosen to self-isolate (resp. shut down) even in the absence of the policy.\footnote{This analysis is, to our knowledge, the first example of an application of Lucas Critique argument in this context; see Section \ref{sec:LucasCritique} for a more articulated discussion and interesting examples.}


\subsection{A general lockdown of the City and School/Work}

We study the effects of a generalized lockdown of the City and 50 percent of School/\-Work locations. We consider policies where the lockdown is applied when the fraction of active cases, $A+Y$, reaches a given fraction of the population. We report results regarding different policies regarding when to reopen the City and the School/Work locations operating remotely. The results reveal the trade-off between flattening the 
the curve of active cases and delaying the reaching
of global herd immunity. 

\subsubsection{Reopen forever after active cases return to  the fraction which triggered the lockdown}

\begin{figure}[!t]
    \centering
    \caption{Lockdown and reopen forever}\label{fig:lockdownonce}
    \includegraphics[height=0.4\textwidth]{images/counterf-lockonce-left.pdf}
    \medskip    
    
	\begin{tabular}{m{0.2\linewidth}
	 				*5{>{\centering}m{0.07\linewidth}}
					>{\centering \arraybackslash}m{0.07\linewidth}}
        \toprule
		& \multicolumn{3}{c}{Steady-state outcomes} & \multicolumn{3}{c}{Share of infections} \tabularnewline 
		 & D & R & Peak A+Y & City & W/S & Home\tabularnewline 
		\midrule
       	(a) No lockdown 	& 0.008         & 0.766         & 0.323         & 0.463         & 0.205         & 0.332      
       	\tabularnewline
       	\midrule 
		(b) 10\% rule 		& 0.007         & 0.703         & 0.158         & 0.363         & 0.225         & 0.412  \tabularnewline 
	    (c) 5.0\% rule		& 0.007         & 0.712         & 0.150         & 0.362         & 0.231         & 0.407 \tabularnewline 
		(d) 3.5\% rule 	& 0.007         & 0.715         & 0.159         & 0.363         & 0.233         & 0.404 \tabularnewline
		\bottomrule 
		\end{tabular}
\bigskip

	\caption*{\footnotesize \normalfont {\em Note:} \tablenote \lineLd.\ \tab.}
\end{figure}

We start by studying the case in which both City and School/Work reopen forever after active infections return to the fraction which triggered the lockdown (10, 5, and 3.5 percent). In this case, because the lockdown may have delayed the population reaching herd immunity, infections may increase again after reopening. 

The results are displayed in Figure \ref{fig:lockdownonce}. We notice that the flattening of the infection curve is substantial. In all cases, the peak of active cases with lockdown is less than half than without lockdown. The reduction of Dead and Recovered at the steady-state is substantial.
Importantly, the lockdown has rather large effects on the Old (not reported in the table). 
The fraction of Dead and Recovered decreases under all policies by about ten percentage points for the Old and only about half as much for the Young (and Not employed). While either lockdown reduces the fraction of Dead at the steady-state of only one-tenth of 1 percent, this reduction is four to five times as big for the Old.   

Notably, the lockdown strategy interacts in rather subtle ways with the dynamics of herd immunity.  When the lockdown occurs too early (5 percent in our simulation), the fraction of infected is too small, herd immunity is too far out, and at reopening a second wave of infections
reaches a  peak higher than the first one.

\begin{figure}[t]
    \centering
    \caption{Cautious reopening}\label{fig:cautious}
    \includegraphics[height=0.4\textwidth]{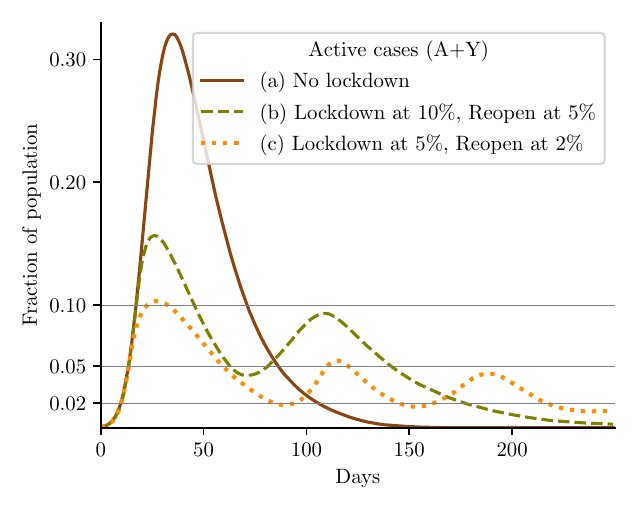}
    \medskip
    
        	\begin{tabular}{m{0.2\linewidth}
	 				*5{>{\centering}m{0.07\linewidth}}
					>{\centering \arraybackslash}m{0.07\linewidth}}
        \toprule
        & \multicolumn{3}{c}{Steady-state outcomes} & \multicolumn{3}{c}{Share of infections} \tabularnewline 
		& D & R & Peak A+Y & City & W/S & Home \tabularnewline 
		\midrule
       	(a) No lockdown 	& 0.008         & 0.766         & 0.323         & 0.463         & 0.205         & 0.332      
       	\tabularnewline
        \midrule 
		(b) 10-5\% rule	& 0.007         & 0.674         & 0.158         & 0.339         & 0.235         & 0.426 \tabularnewline 
		(c) 5-2\% rule	& 0.006         & 0.630         & 0.104         & 0.292         & 0.254         & 0.453 \tabularnewline 

		\bottomrule
		\end{tabular}
		\bigskip

    	\caption*{\footnotesize \normalfont {\em Note:} 
        \tablenote \lineLdReop.\  \tab.   	}
\end{figure}

\subsubsection{Cautious reopening}\label{sec:cautious}
We study rules in which reopening is set more cautiously: a 5 percent reopening threshold for the 10 percent lockdown and a 2 percent reopening threshold for the 5 percent lockdown. This is an interesting health policy strategy because, as we have seen in the previous sections,  the lockdown strategy interacts with herd immunity dynamics, and an early reopening can induce a high second wave of infections.

Results are in Figure  \ref{fig:cautious}. Relative to the case in which reopening occurs at the same threshold as the lockdown, herd behavior is more advanced at reopening, and the second wave is substantially dampened. The peaks of the second wave occur at about 8 percent as opposed to 13 percent of the population, for the 10 percent lockdown. The result is even more striking for the 5 percent lockdown. In this case, with reopening at 5 percent, the second wave was higher than the first (15 percent). With cautious reopening at 2 percent, instead, the second wave is dampened to 5 percent. Interestingly, however this triggers another lockdown and hence a third wave whose peak is below 5 percent. Relatedly, the fraction of Dead and Recovered at steady-state is lower with cautious reopening, about 3 and 8 percentage points, respectively, for the 10 and the 5 percent lockdowns. Cautious reopening, we conclude, is particularly effective for the 5 percent lockdown, where reopening at 5 percent induced a large second wave and a fraction of Dead and Recovered at steady-state larger than with a less restrive lockdown at 10 percent (with reopening at 10 percent). 

\subsection{Local (Neighborhood) Lockdowns} 

In this section we allow public health authorities to define  lockdown rules depending on local neighborhood-level information about the epidemic. The aim is to investigate whether such granular information is relevant  to improve the effectiveness of to NPI policies. 

\begin{figure}[t]
    \centering
    \caption{Neighborhood-specific lockdowns - Different number of neighborhoods\label{fig:nspec}}

        \includegraphics[height=0.4\textwidth]{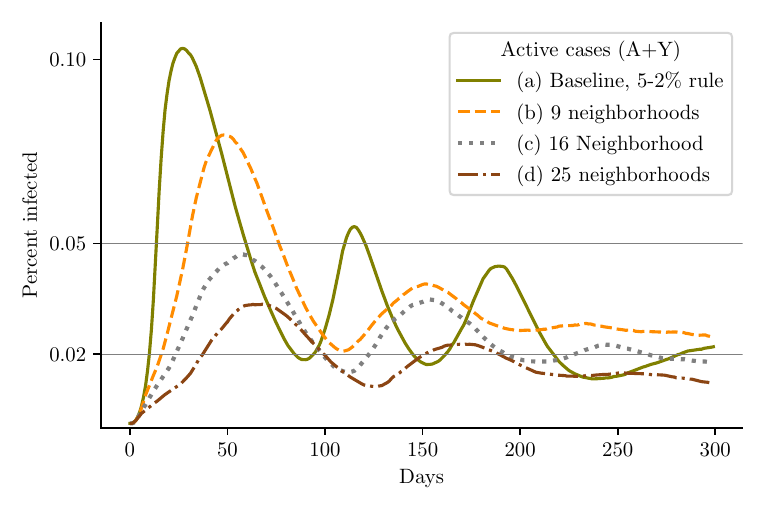}
\medskip
        	\begin{tabular}{m{0.24\linewidth}
	 				*5{>{\centering}m{0.07\linewidth}}
					>{\centering \arraybackslash}m{0.07\linewidth}}
        \toprule
        & \multicolumn{3}{c}{Steady-state outcomes} & \multicolumn{3}{c}{Share of infections} \tabularnewline 
		& D & R & Peak A+Y & City & W/S & Home \tabularnewline 
		\midrule
		(a) Baseline 5-2\% rule 		& 0.006         & 0.630         & 0.104         & 0.292         & 0.254         & 0.453 \tabularnewline 
	\midrule
		(b) 9 neighborhoods 	 & 0.006         & 0.631         & 0.087         & 0.285         & 0.262         & 0.453  \tabularnewline 
		(c) 16 neighborhoods 	& 0.004         & 0.462         & 0.065         & 0.257         & 0.286         & 0.457 \tabularnewline 
		(d) 25 neighborhoods 	& 0.003         & 0.335         & 0.047         & 0.251         & 0.288         & 0.461 \tabularnewline 
	 \bottomrule
	 \end{tabular} 

    	\caption*{\footnotesize \normalfont {\em Note:} 
        The figure illustrates the dynamics of active cases ($A+Y$) as a fraction of the total population under a cautious reopening policy (5-2\% rule) applied at the neighborhood level. The green solid line (Baseline) corresponds to a policy applied city-wide. \lineLdReop. \tab.   	}


\end{figure}


\begin{figure}[t]
    \centering
    \caption{Neighborhood-specific lockdowns - School/Work Mixing\label{fig:nnspec}}
    
    \begin{subfigure}{.5\textwidth}
        \centering
       \includegraphics[width=\textwidth]{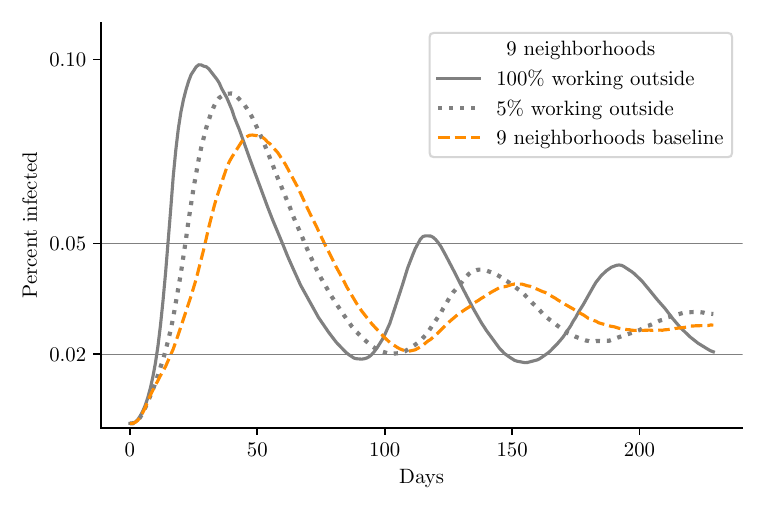}
        \caption{9 Neighborhoods}\label{fig:firmmix}
    \end{subfigure}%
   \begin{subfigure}{.5\textwidth}
        \centering
            \includegraphics[width=\textwidth]{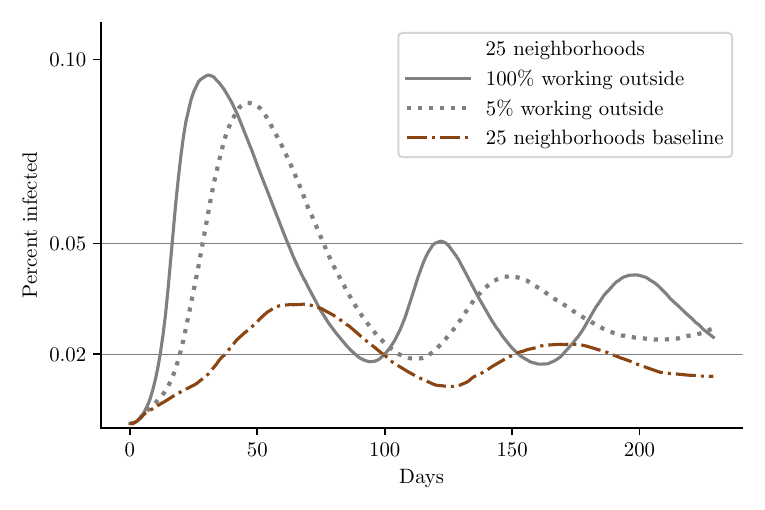}
    \caption{25 Neighborhoods}
    \label{fig:my_label}
      \end{subfigure}%
    	\caption*{\footnotesize \normalfont {\em Note:} 
        The figure illustrates the dynamics of active cases ($A+Y$) as a fraction of the total population under a cautious reopening policy (5-2\% rule) applied at the neighborhood level, under different levels of School/Work mixing across neighborhoods. The orange dashed line in panel (a) and the brown dash-dotted line in panel (b) correspond to the lines in Figure \ref{fig:nspec}. \lineLdReop. }

\end{figure}
Specifically,  we divide the city in {$S={9,16,25}$} symmetric square-shaped neighborhoods. Home and  School/Work are neighborhood specific and agents residing in a given neighborhood visit a School/Work  in that same neighborhood. Agents visit the City initially in a random location of the neighborhood where they reside, but we let them move across neighborhood borders. The initial cluster of infection is in the south-west corner of the city. We simulate the effect of the cautious lockdown and reopening policy described in Section \ref{sec:cautious}, but with thresholds and lockdowns that are neighborhood-specific.  Results are shown in Figure \ref{fig:nspec}: neighborhood-specific NPI policies are indeed very effective, as - other things equal - they induce substantial flattening of the epidemic  infections and a smaller number of total infected and fatalities at steady state. Not surprisingly, but still importantly, this effect is monotonic in the number of neighborhoods, indicating the importance on the part of public health authorities of collecting and leveraging in policy design  more and more granular information about infections.   

In Figure \ref{fig:nnspec} we allow a fraction of agents to attend School/Work  outside of the neighborhood where their Home is located and they are initially placed. In both the 9 and 25 neighborhood simulations - respectively in Figures \ref{fig:firmmix} and  \ref{fig:my_label} - mixing of School-Work has a very large negative effect on the ability of NPI to  flatten of the epidemic. Indeed, in both cases just a $5\%$ of student/workers outside the neiborhood is enough to increase dramatically the peak of infections, keeping NPI policies unvaried. 

\subsubsection{Heterogeneous neighborhoods}

To investigate the role of neighborhood heterogeneity in the dynamics of an epidemic, we simulate  a city divided in 9 neighborhoods as in the orange line of Figure \ref{fig:nspec}, but we place Homes with larger households - which we can interpret as poorer - closer to the southwest corner of the city (neighborhood 1, with average household size equal to 8.3), and progressively smaller household as the location of residence gets closer to the northeastern corner, where only singles live (neighborhood 9, with average household size about 1). We however maintain neighborhoods density constant across all neighborhoods, to isolate the effect of household size from the effect of density we studied in \cite{bisinmoro2020Geo}, and we report simulations where School/Work locations for all agents are in their  neighborood of residence.

\begin{figure}[t]
    \centering
    \caption{Neighborhoods with different household size}\label{fig:hetneigh}
    \includegraphics[height=0.4\textwidth]{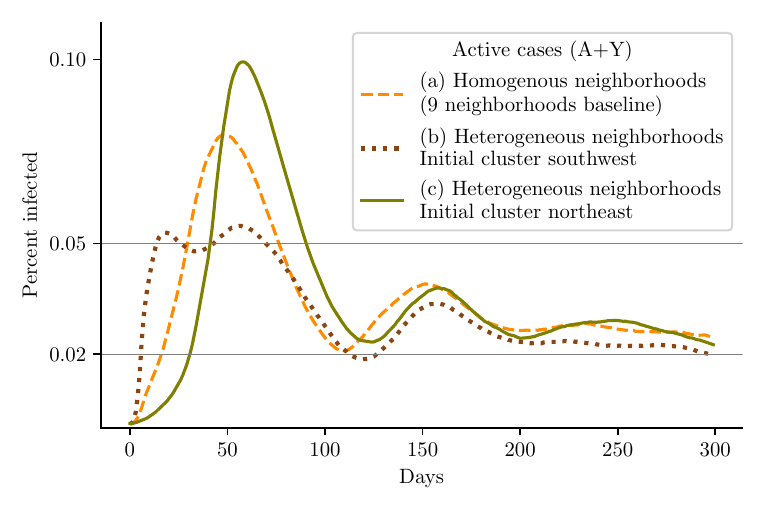}
    \medskip
    
	 \begin{tabular}{m{0.33\linewidth}
	    *5{>{\centering}m{0.07\linewidth}}
	    >{\centering \arraybackslash}m{0.07\linewidth}}
	\toprule 
& \multicolumn{3}{c}{Steady-state outcomes} & \multicolumn{3}{c}{Share of infections} \tabularnewline 
& D & R & Peak A+Y & City & W/S & Home \tabularnewline 
	\midrule 
		(a) Homogenous neighb. 	& 0.006         & 0.631         & 0.087         & 0.285         & 0.262         & 0.453 \tabularnewline 
	\midrule
		(b) Heterogeneous, SW cluster 	 & 0.006         & 0.562         & 0.067         & 0.283         & 0.267         & 0.450 \tabularnewline 
		(c) Heterogeneous, NE cluster 	 & 0.006         & 0.639         & 0.104         & 0.307         & 0.248         & 0.446 \tabularnewline
	 \bottomrule
	 \end{tabular} 
    \medskip
    
    \caption*{\footnotesize \normalfont {\em Note:} 
        The figure illustrates the dynamics of active cases ($A+Y$) as a fraction of the total population under a cautious reopening policy (5-2\% rule) in a city with 9 neighborhoods differing by household size. The dashed yellow line corresponds the 9-neighborhoods line in Figure \ref{fig:nspec}.  \tab.}
\end{figure}

\begin{figure}[!t]
    \centering
    \caption{Dynamics by neighborhood}\label{puntini}
\label{hetneighbyneigh}
        \begin{subfigure}{.5\textwidth}
        \centering
        \includegraphics[width=.7\textwidth]{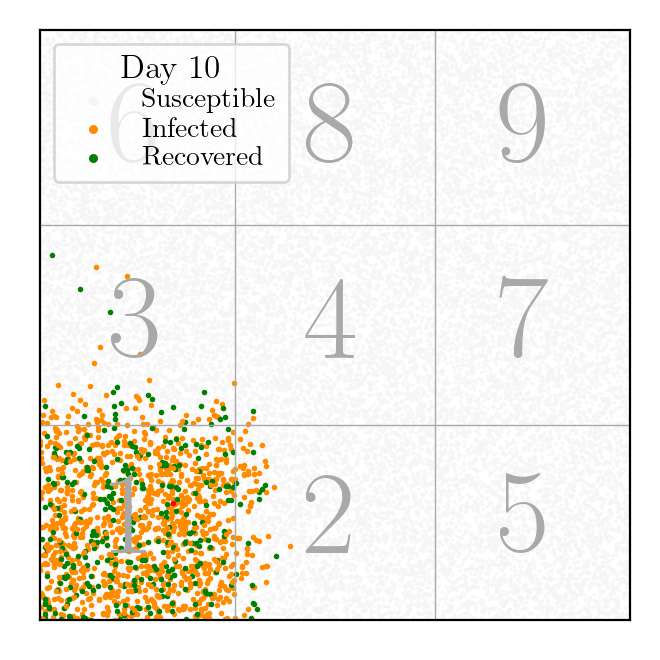}
    \end{subfigure}%
    \begin{subfigure}{.5\textwidth}
        \centering
        \includegraphics[width=.7\textwidth]{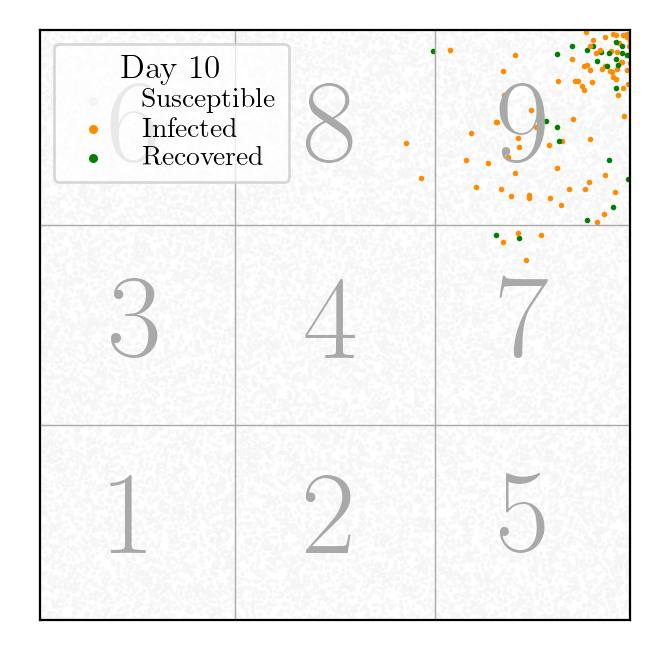}
    \end{subfigure}
    
    \begin{subfigure}{.5\textwidth}
        \centering
        \includegraphics[width=\textwidth]{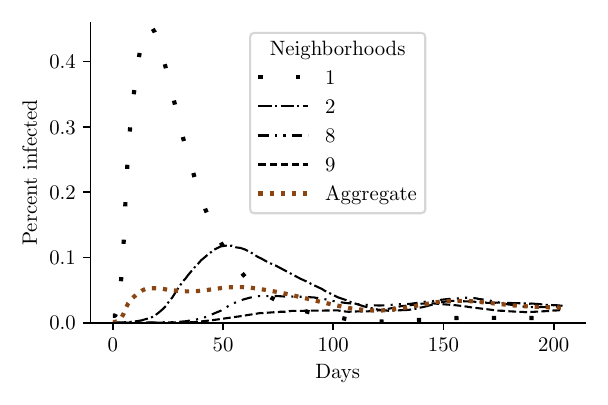}
        \caption{Initial cluster near highest average household size (southwest)}\label{fig:het-byn}
    \end{subfigure}%
    \begin{subfigure}{.5\textwidth}
        \centering
        \includegraphics[width=\textwidth]{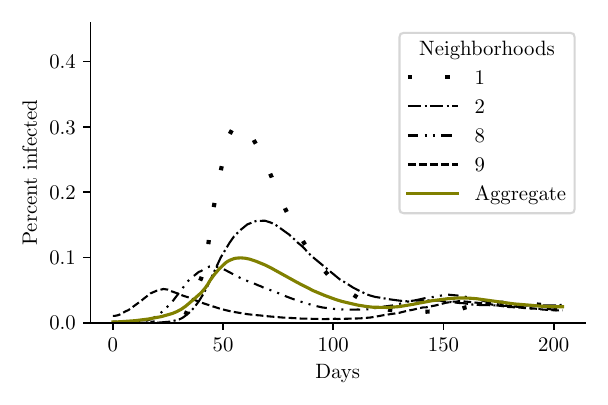}
        \caption{Initial cluster near lowest average households size (northeast)}\label{fig:het-rev-byn}
    \end{subfigure}

\caption*{\footnotesize \normalfont {\em Note:} 
        The top two panels display the geographic location of individuals color-coded by their status at day 10. The bottom panels display the dynamics of active cases ($A+Y$) as a fraction of the total population in selected neighborhoods under a cautious reopening policy (5-2\% rule) in a city with 9 neighborhoods differing by household size. Neighborhood 1 is located in the southwest corner of the city and is the neighborhood with highest average family size. Neighborhoods 2 \ldots8 are located progressively further away from neighborhood 1 and include households with progressively smaller family size. Neighborhood 9 is located in the northeast corner and is the neighborhood with the lowest average family size. The brown dotted line in panel (a) and the green solid line in panel (b) correspond to the aggregate behavior in Figure \ref{fig:hetneigh}.}
\end{figure}

		

In Figure \ref{fig:hetneigh} we show that the dynamics of infections  spreading from a cluster near the southwest corner (with largest average household size) display an earlier but  lower peak   than the corresponding dynamics starting from the northeast corner (with smallest average household size). This is an interesting composition effect of the heterogeneity in average household size. Figure \ref{fig:het-byn} show that an infection spreading from the neighborhood with the largest average household size (neighborhood 1) has an early and large peak in this neighborhood, which spreads into into the other neighborhoods but dies off quickly, inducing local herd  immunity at the level of the neighborhood before spreading to lower average household size neighborhoods (like 8 and 9). On the other hand, an infection spreading from the neighborhood with the lowest average household size (neighborhood 9) in Figure \ref{fig:het-rev-byn} starts slowly because this neighborhood includes only singles, and no-one can be infected at home. The infection builds up to induce relatively high peaks of infections at the same time in the neighborhood with mid-average household size, inducing an higher aggregate peak of infections. The different speed at which the infection spread in the two cases - when the cluster starts from the highest or the lowest average household size - is clearly shown comparing the state of the dynamics of infection after 10 days, as in the top two panels of Figure \ref{puntini}. The differences are dramatic even though, it is important to remark, the densities of the neighborhoods are maintained constant.

\subsection{Selective Lockdowns}

This section presents two selective lockdown policies. In the first one, we limit the lockdown to the City, and we let firms/schools decide whether to operate remotely. In the second one, we limit the lockdown to the Old, a policy aiming at reducing total fatalities while bearing limited economic costs (the Old are not economically active but suffer a higher fatality rate than other demographic groups). 
Both these lockdown policies have the important property that they are much less costly, economically, as worker/students remain active. This consideration needs to be traded-off with additional costs in terms of health outcomes at the steady-state or at the peak, e.g. if the society has a constraint with respect to health care resources like hospital/ICU beds. Furthermore, while these selective lockdowns have positive direct effects in terms of economic costs,  indirect effects across locations and/or demographic types could, in principle, substantially limit their advantage over general lockdowns. 

\subsubsection{City-Only lockdown}

\begin{figure}[t]
    \centering
    \caption{City-only lockdown}\label{fig:selectivecity}
    \includegraphics[width=0.55\textwidth]{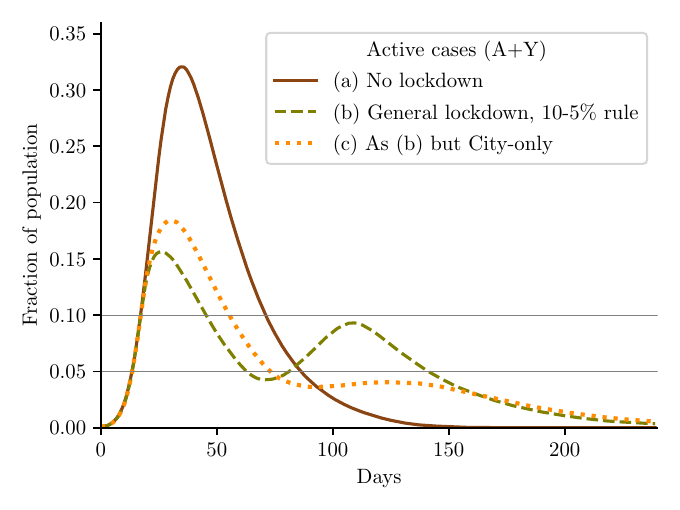}

    \medskip
    
    \begin{tabular}{m{0.3\linewidth}
 		*5{>{\centering}m{0.07\linewidth}}
		>{\centering \arraybackslash}m{0.07\linewidth}}
		\toprule 
		& \multicolumn{3}{c}{Steady-state outcomes} & \multicolumn{3}{c}{Share of infections} \tabularnewline 
		& D & R & Peak A+Y & City & W/S & Home  \tabularnewline 
		\midrule 
		  (a) No lockdown 	& 0.008         & 0.766         & 0.323         & 0.463         & 0.205         & 0.332      
       	\tabularnewline
       	\midrule
		(b) Baseline 10-5\% rule	& 0.007 	& 0.674 	& 0.158 & 0.339 	& 0.235 	& 0.426 	\tabularnewline 

		 (c) City-only lockdown 	  & 0.006         & 0.650         & 0.186         & 0.264         & 0.296         & 0.440 \tabularnewline 
		\bottomrule

		\smallskip
		\end{tabular}

		\caption*{\normalfont \footnotesize {\em Note:}
		\tablenote \lineLdReop. \tab.}
\end{figure}

Results for the City-Only lockdown are reported in Figure \ref{fig:selectivecity}, with respect to a 10 percent lockdown with cautious reopening at 5 percent. Interestingly, in this case, with respect to the general lockdown, the first peak in $A+Y$ is higher (18.6 percent rather than 15.8), but there is no significant second wave. Furthermore, the total fraction of Dead and Recovered at the steady-state is two percentage points lower. 
This is striking, as the City-Only lockdown is much less costly, economically, than the general lockdown. According to these simulations, the City-Only lockdown has extra health costs inasmuch as the first wave might be too high with respect to a possible constraint in terms of health care resources. In other words, indirect effects, e.g., from City to Home, do not appear to limit the advantages of the City-Only lockdown in our simulations.\footnote{As we repeatedly noted, we do not aspire to quantitatively precise prediction because the calibrated model we simulate is very stylized. }

\subsubsection{Old-Only lockdown}

\begin{figure}[!t]
    \centering
    \caption{Old-only lockdown}    \label{fig:selectiveold}
    \begin{subfigure}{.5\textwidth}
    \centering
    \includegraphics[width=\textwidth]{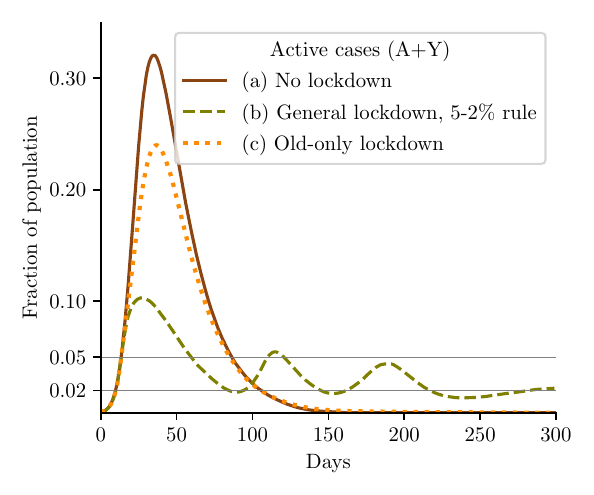}
    \caption{Old-only Lockdown at home} 
    \end{subfigure}%
        \centering
        \begin{subfigure}{.5\textwidth}
        \includegraphics[width=0.83\textwidth]{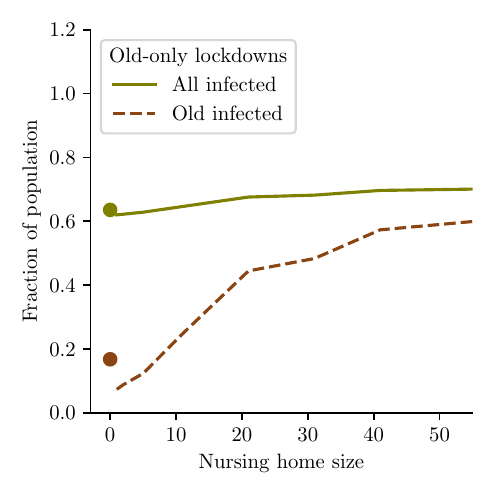}
        \caption{Lockdown in nursing homes, by size}\label{fig:counterf-nursinghomes}
    \end{subfigure}
    \medskip
    
    \begin{tabular}{m{.02\linewidth}>{\raggedright}m{0.26\linewidth}
	 	*5{>{\centering}m{0.07\linewidth}}
        >{\centering \arraybackslash}m{0.07\linewidth}}
		\toprule 
		&& \multicolumn{3}{c}{Steady-state outcomes} & \multicolumn{3}{c}{Share of infections} \tabularnewline 
		&& D & R & Peak A+Y & City & W/S & Home \tabularnewline 
		\midrule 
       	\multirow{2}{*}{(a)} &  All, no lockdown 	& 0.008 	& 0.766 	& 0.323  & 0.463 	& 0.205 	& 0.332 	 
       	    \tabularnewline
        & Old, no lockdown 		& 0.029 	& 0.599 	& 0.249 & 0.726 	& - 	& 0.274 	\tabularnewline

       	\midrule
 		 (b) & Old, General lockdown 		& 0.019 	& 0.397 	& 0.050 & 0.609 	& - 	& 0.391 	\tabularnewline 
        \midrule
		 \multirow{2}{*}{(c)} & All, Old-only lockdown	& 0.004 	& 0.635 	& 0.241 & 0.436 	& 0.149 	& 0.415  	  \tabularnewline 
 		& Old, Old-only lockdown & 0.008 	& 0.167 	& 0.053 & 0.228 	& - 	& 0.772 	\tabularnewline 
		\bottomrule

		\smallskip
		\end{tabular}

		\caption*{\normalfont \footnotesize {\em Note:} The figure on the left and the table illustrate, respectively, the dynamics of active cases ($A+Y$) and the steady-state outcomes, under a cautious lockdown policy (5-2\% rule) when Old are isolated at their home. 
		The figure on the right illustrates the frations of infected among all and Old (as a fraction of the population) when the Old are isolated in nursing homes of different size (the dots represent outcomes when Old are isolated at their homes.}
\end{figure}

Results for the Old-Only lockdown are reported in Figure \ref{fig:selectiveold}, simulating a 5 percent lockdown rule with a period of cautious reopening at 2 percent. Olds are isolated in their Homes, where they are in contact with their families. 
Comparing it with the case in which the lockdown is general (that is, includes the Young and the Not Employed as well), it is remarkable that the number of stationary Dead between the Old is more than halved, from almost 2 percent to less that 1 percent. The peak of active cases remains rather high, as the  Young and the Not Employed keep interacting in the City and the Young also at School/Work. This high peak does not necessarily affect possible constraints on health care resources if the Old use them relatively more intensely. 
As for the City-Only lockdown, our simulations of the Old-Only lockdown suggest that indirect effects, e.g., from Young to Old or from School/Work to Home, do not substantially limit its advantage. 

The simulations we have just reported refer to the case in which the Old are locked-down at Home, where they interact with Young and Not Employed agents. In Figured \ref{fig:counterf-nursinghomes} we report $D+R$ at steady-state for different simulations where the Old are locked-down in nursing homes of different sizes (the dot is the case where the Old are locked-down at their Home). It is noteworthy that the number of Dead increases steeply with the size of the nursing homes, reaching 2 percent for homes with 50-agents each. It is also interesting that the lockdown of the Old at Home is equivalent in terms of infection and fatality rates to one where the Old are locked-down in nursing homes of about size 10. This is a consequence of the indirect interactions across locations hurting the more susceptible demographic group.  


\section{An Exercise in the Lucas Critique \label{sec:LucasCritique}}

This section illustrates the implications of what economists refer to as the {\em Lucas Critique}, in the context of epidemiological models.\footnote{From \citet{lucas1976econometric}. In its original formulation, the Critique is summarized as follows: 
\begin{quotation} ``Given that the structure of an econometric model consists of optimal decision rules of economic agents and that optimal decision rules vary systematically with changes in the structure of series relevant to the decision-maker, it follows that any change in policy will systematically alter the structure of econometric models.'' \end{quotation} }
We have already shown, in the previous section, that the behavioral response of agents and firms to the dynamics of an epidemic is bound to have important effects on the dynamics themselves. We now discuss how policy evaluations and interventions disregarding that agents' and firms' ``decision rules vary systematically with changes in the structure of series relevant to the decision-maker'' might lead to policy decisions that are very costly in terms of their effects on the dynamics of the epidemic. 
\subsection{Policymaker error}
\label{sec:lock-nobeh}

\begin{figure}[!t]
    \centering
    \caption{Naive policymaker}

    \begin{subfigure}{.5\textwidth}
        \centering
        \includegraphics[height=0.8\textwidth]{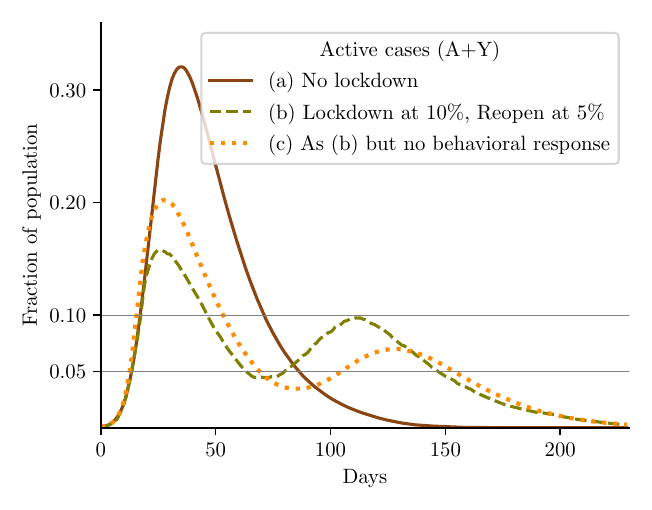}
        \caption{Cautious reopening, 10-5\% rule}\label{fig:counterf-lock-nobeh}
    \end{subfigure}%
    \begin{subfigure}{.5\textwidth}
        \centering
        \includegraphics[height=0.8\textwidth]{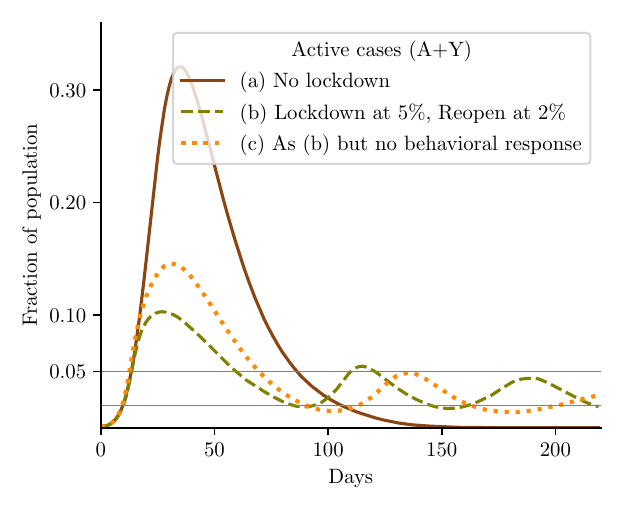}
        \caption{Cautious reopening, 5-2\% rule}\label{fig:counterf-lock-5-2-nobeh}
    \end{subfigure}%

		\caption*{\normalfont \footnotesize {\em Note:}
		\tablenoteLC \lineLdReop.}
\end{figure}

We simulate the prediction error of a policymaker evaluating different lockdown policies while not anticipating the behavioral response of agents and firms to the dynamics of the epidemic. 
 We interpret the dynamics without behavioral response as the policymaker's prediction of the effect of the lockdown and the case with response as the actual outcome of the lockdown. 

We report the case of a lockdown at 10 percent with reopening at 5 and a lockdown at 5 percent with reopening at 2; respectively in Figure \ref{fig:counterf-lock-nobeh} and \ref{fig:counterf-lock-5-2-nobeh} (we do not report tables for these simulations). In both cases, the policymaker predicts a higher peak than it actually occurs. On the other hand, he/she will face an earlier and higher than predicted second wave.

\subsection{Policymaker error when recalibrating the reopening policy }

In this section we postulate a more articulate procedure to evaluate different lockdown policies on the part of a policymaker who does not anticipate the behavioral response of agents and firms. 
Consider a policymaker who evaluates lockdown interventions assuming no behavioral responses, but chooses the reopening threshold as a consequence of a policy re-valuation after the lockdown, still disregarding the behavioral response. 
More specifically, consider the case where, at the re-evaluation stage, the policymaker naively attributes the lower-than-predicted peak of infections to the effect
of fewer people visiting the City and School/Work locations, generating lower contact
rates and slower diffusion of the epidemics. The policymaker, however, continues to disregard the fact that behavioral responses are a function of the number of cases and re-evaluates the reopening policy assuming that
people visiting the City and School/Work locations remain at the (lower) level observed at the re-evaluation. 

\begin{figure}[!t]
    \centering
    \caption{Re-calibrated reopenings}
    \begin{subfigure}{.5\textwidth}
        \centering
        \includegraphics[height=0.8\textwidth]{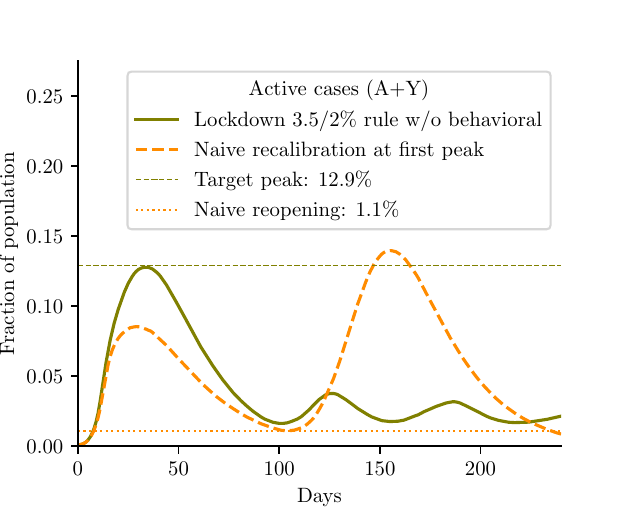}
        \caption{Initial rule: 3.5\% lockdown}\label{fig:lcr2}
    \end{subfigure}%
    \begin{subfigure}{.5\textwidth}
        \centering
        \includegraphics[height=0.8\textwidth]{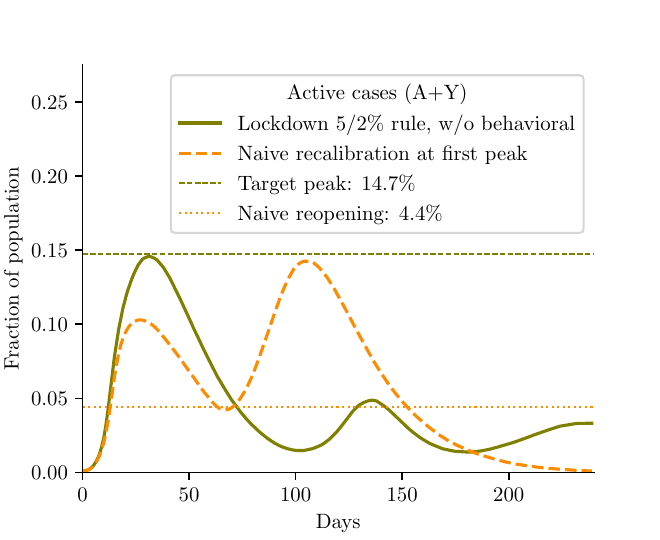}
        \caption{Initial rule: 5\% lockdown}\label{fig:lcr1}
    \end{subfigure}%

		\caption*{\normalfont \footnotesize {\em Note:}
		\tablenoteLC The dashed horizontal lines indicate the initial (naive) target peak. The dash-dotted horizontal lines indicate the recalibrated reopening. Panel (a) Lockdown threshold policy at 3.5\% with a target peak at 12.9 (dashed line). Results from adjusting the reopening policy at 1.1\% (dotted line) after the first peak. Panel (b): Lockdown threshold policy at 5\% with a target peak at 14.7\% (dashed line). Results from adjusting the reopening policy at 4.2\% after the first pea (dotted line).}
		
\end{figure}

The adverse effects and the costs of this procedure are clearly seen in the  
case in which  a policymaker sets up a 3.5 percent lockdown rule with the goal
of not experiencing a peak active cases over about 13 percent, as predicted by the model without behavioral response (see Figure \ref{fig:lcr2}). When reaching the peak, which occurs at about 8 percent, the policymaker re-evaluates its reopening policy, re-calibrates its model and computes a new reopening rule. The recalibration and the new reopening rule are computed postulating still no behavioral response upon reopening. In our simulations, we run the model to find
the reopening rule that would generate 13 percent active cases at
the peak in the policymaker's recalibrated model.  

The error induced by the policymaker recalibrating the reopening policy is evident in the Figure. 
Because there is not enough herd immunity at the first peak,  the policymaker delays reopening until the fraction of active cases is about 1 percent. 
Nonetheless, the recalibration misses the behavioral response of agents and firms, who increase the contact rate in both the City and the School/Work due to the low infection rates. As a consequence,  the second wave is much higher than the first and ends up reaching a peak of 14 percent, 9\% higher than the government's original target. If the lockdown and opening threshold are calibrated to avoid hitting a  constraint on the available health care resources, this simulation implies that health care resources will be under-utilized after lockdown (that is, the lockdown will turn out to be stricter than necessary) and they will instead be over-run after reopening, with potentially high human costs.  

Interestingly, the costs associated with this policy-making procedure disregarding behavioral responses are less apparent in the case of a lockdown at 5 percent so as to avoid a peak higher than about 15 percent of active cases (see Figure \ref{fig:lcr1}). It is still the case that, after observing a smaller than expected first peak, the policymaker, expects a low infection rate for the future, induced by (what she expects to be a constantly)  lower contact rate. She then decides to reopen too soon, when cases are low, and hence the behavioral response of agents and firms increases the contact rate. As a consequence, the epidemic picks up, giving rise to an unexpected second wave of infection as agents and firms. 
However, in this case, the second peak after recalibration is higher than the first, but near the 15 percent target, not higher as in the previous case. While the recalibration misses the behavioral response, leading agents and firms to increase the contact rate after the first peak, it also misses the herd immunity generated by this behavior. The effect of herd immunity is larger because initial lockdown is set at 5 percent rather than at 3.5 percent, because of the relatively higher first peak of infections, and because, after the government re-evaluation, when active cases decline more
people than the government expects visit the City and get infected.

\begin{figure}[t]   \centering
    \caption{Lucas Critique with Neighborhood-Specific Lockdowns}
   \begin{subfigure}{.5\textwidth}
        \centering
    \includegraphics[width=\textwidth]{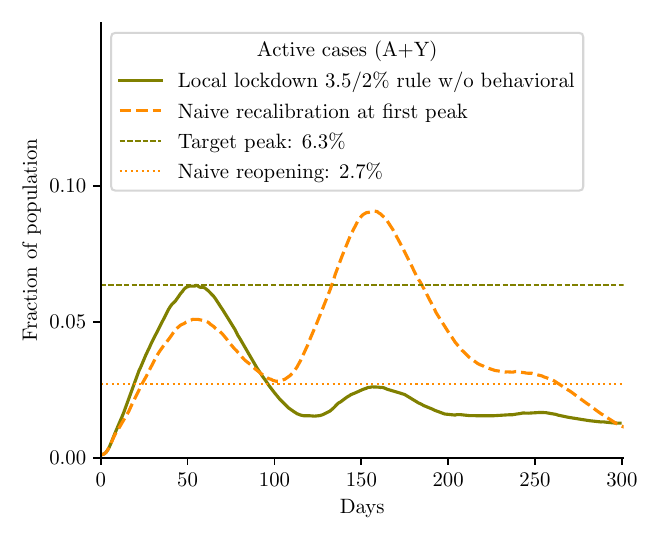}
    \caption{Homogeneous neighborhoods}\label{LCN} 
    \end{subfigure}%
       \begin{subfigure}{.5\textwidth}
        \centering
    \includegraphics[width=\textwidth]{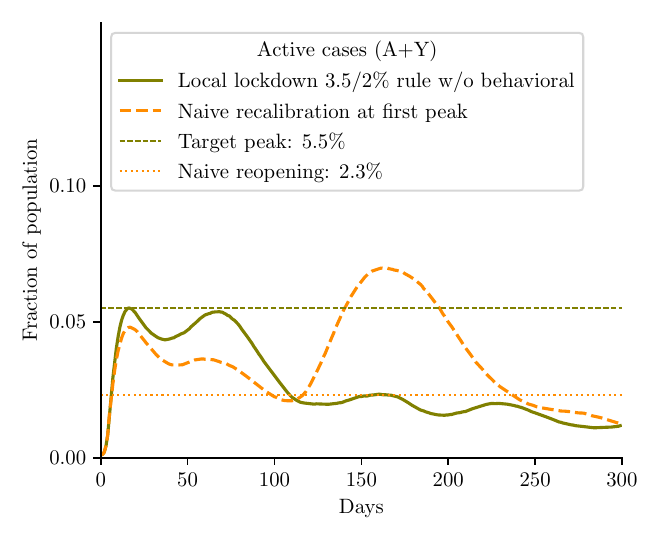}
    \caption{Heterogeneous neighborhoods}\label{fig:LCNhet}
    \end{subfigure}
\medskip
		\caption*{\normalfont \footnotesize {\em Note:}
		\tablenoteLC The dashed horizontal lines indicate the initial (naive) target peak. The dotted horizontal lines indicate the recalibrated reopening. Panel (a): model with 9 homogenous neighborhoods; Panel (b) model with 9 heterogeneous neighborhoods (initial cluster in the soutwhest courner).} 
\medskip
\end{figure}

The policymaker error when naively recalibrating the reopening policy in an environment where lockdowns are neighborood-specific illustrates another interesting dimension of application of the Lucas critique to NPI policies. In the model with 9 homogeneous neighborhoods and neighborood-specific lockdown policy, the naive reopening, after naively recalibrating the model, induces a second peak overshooting the target of 6.3\%, computed with the same criteria used in the previous cases, by over 40\% of its level, reaching a second peak of infection at about 9\% infected, as shown in Figure \ref{LCN}.

A qualitatively similar outcome is obtained if we repeat the exercise with heterogeneous neigborhoods differing by average household size, with the initial outbreak in the corner of the city with the largest household size (the same structure used in the simulations reported in Figure \ref{fig:het-byn}). The peak after the government's recalibration overshoots the target by 26\% (see Figure \ref{fig:LCNhet}).\footnote{When the outbreak starts in the neighborhood with the smallest average household size, the recalibration does not result in overshooting.}

\section{Conclusions} 

In this paper, we developed a parsimonious stylized model which, in our objectives, adds several dimensions  of interest to the standard epidemiological components of general economic models in which the policy trade-off between economic activity and epidemic diffusion is evaluated, as, e.g., in \cite{atkeson2020will},  \cite{alvarez2020simple},  \cite{jarosch2020internal}, \cite{kaplan2020gianluca}. Both the geographic and the network dimension appear very relevant in several empirical analyses of the effects of the SARS-CoV-2 epidemic, which tend to concentrate on economic activity granular level.\footnote{See e.g., along these lines, \cite{alstadsaeter2020first},  \cite{chetty2020economic}, \cite{crossley2020heterogeneous}, \cite{note2020spending},  and \cite{de2020preparing}.} 
The model is designed to qualitatively and quantitatively identify the role of the interactions between several fundamental ``forces'' driving the dynamics of an epidemic, like  (global and local) herd immunity, network, and demographic heterogeneity, behavioral responses of agents and firms to infections, and Non-Pharmaceutical policy interventions. In this respect, we stress that the role of Lucas Critique arguments, which identify the adverse effects and the costs of policy interventions disregarding behavioral responses, is potentially first order, qualitatively and quantitatively, depending on the calibration. 

\newpage
\appendix

\section{Appendix: Theoretical Structure of SIR and Spatial-SIR } \label{sec:app_theory}
We refer to Appendix A in \cite{bisinmoro2020Geo} for a formalization of the Spatial-SIR we extend in this paper. The extension we adopt in this paper adds demographic and network structure as described but does not alter the mechanics of the epidemic state transitions. 

\section{Appendix: Calibration} \label{app:calibration}

We calibrate the parameters of Spatial-SIR  to the dynamics of the SARS-CoV-2 epidemic. We acknowledge the substantial uncertainty in the literature with respect to even the main epidemiological parameters  pertaining to this epidemic. As we noted in the introduction, this is less problematic when aiming at understanding mechanisms and orders-of-magnitude rather than at precise forecasts. The parameters we choose in the calibration of the aggregate model
are summarily reported in Table \ref{parameters}. 

\subsubsection*{Geography}

The City is a square of area equal to 1. We choose $N=25600$ so that our simulations converge in a reasonable time (see \cite{bisinmoro2020Geo} for a description of how the model scales in size). People are located in the City by randomly
drawing their $x$ and $y$ coordinates indipendently from a Uniform distribution $\sim U[0,1]$. At all $t>0$, individuals 
are relocated at distance $\mu$ from their location at $t-1$, in a direction
drawn randomly from a Uniform distribution  $\sim U[0,2\pi]$. School/Work locations are
are squares of size calibrated as described
below. The location of individuals within School/Work locations is randomly drawn as in the City, but does not change over time. Homes have no dimension. All School/Work and Home locations are far
from each other and far from the City
to prevent contagion across different location types, across different schools/workplaces and across different homes. 

We calibrate the relative density of the three locations to match the data on the distribution of contacts reported by \citet{Mossong_2008}.\footnote{They
survey mixing patterns on a sample of the population in eight European
countries. We do not exploit cross-country heterogeneity in this section and consider averages.}
In this data, on average, people make 13.5 contacts per
day. Furthermore, the total number of contacts is distributed as follows: 34 percent at work or at school, 34 percent in other locations besides home.\footnote{We ignore the 7 percent of contacts occurring in ``multiple'' locations.} Assuming that interactions at Home necessarily induce contacts between family members, and with an  average family size of  3.3 from the ACS data,  this implies 2.3 contacts at home.\footnote{This is lower than implied by  \citet{Mossong_2008}, but it's an upper bound for the size distribution of families form ACS we use. }
Of the remaining 11.2 contacts, according to the data in  \citet{Mossong_2008}, about half occur at Work and School and half in what we have categorized as the City. Because a fraction of the population is structurally restricted  not to visit the School/Work location (the Old and the Not employed), we need to rescale the number of contacts per day in each location. In the ACS this fraction is 15 percent and therefore we calibrate that a young and  occupied agent (a schoolchild or a worker) receives 17.6 percent (100/85 percent) more contacts that anyone in other locations. We allocate therefore
the 11.2 contacts per day not occurring at Home proportionally, 6 at School/Work and 5.2 in the City. 
We therefore calibrate the contagion radius to match 5.2 contacts on average in the City, given its population. We keep the same contagion radius in the School/Work locations, 
and calibrate the size of such locations so as to generate 6 contacts, on average.\footnote{With a population of 26,600 individuals on a square of unit side,  the contagion radius we obtain is 0.00805. With 100 agents at each School/Work location on a square, the side we obtain is 0.0547.}

\subsubsection*{Agent types and family composition}
We use data from the 2018 American Community Survey (ACS) to classify agents by type and
allocate them to households replicating the size and age distribution of 
American households. Specifically, we calculate from the ACS the distribution of family compositions by Young/Old age of respondents not leaving in Group Quarters, and construct households of sizes and Young/Old compositions replicating
this distribution. The age cutoff for being classified as Old is 65. We then create households of size 50 containing either only Young or only Old individuals to replicate the fraction of Young and Old ACS respondents living in group quarters (2.9 percent of Old and 2.4 percent of Young). 
In the 2018 ACS, approximately 16 percent of individuals are above 65 years of age. We divide the remaining population to reflect the proportion of the population that are young and either employed or in school, 65 percent, and define the rest, 19 percent,  as Not employed. We randomly assign all Young to a number of School/Work locations such that their average size is 100 individuals. 

%
%

\subsubsection*{Infection and contact rates}
\begin{figure}
    \centering
    \caption{Growth rates of infections in simulation, fatalities in Lombardy, Italy}
    \label{fig:growths-1}
    \includegraphics[width=3in]{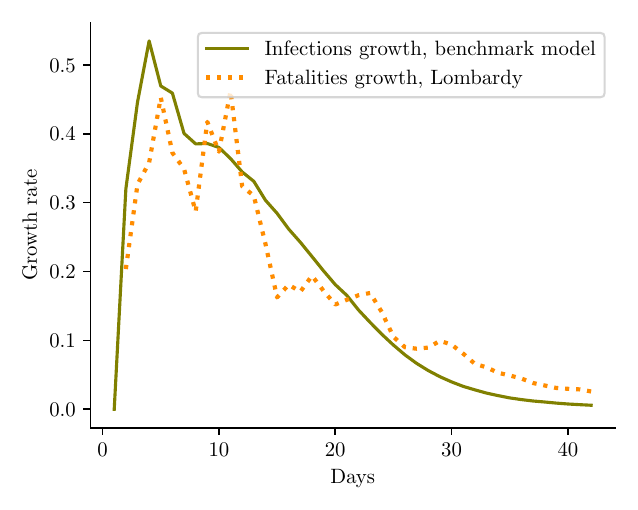}
    \caption*{\footnotesize\normalfont \textit{Note:} Solid green line: average of 20 simulations of the calibrated benchmark model. Dotted yellow line: COVID-19 fatalities growth rate in the Lombardy region (Italy), 3-day moving average, starting from February 29, 2020. Data from the Italian government, Dipartimento di Protezione Civile, available at \href{https://github.com/pcm-dpc/COVID-19}{https://github.com/pcm-dpc/COVID-19}.}
\end{figure}
We calibrate the contagion rate $\pi$ and the mean travel distance $\mu$ to match the daily
growth rates of the dynamics of fatalities, $g_t$, observed in the first 35 days of epidemics in Lombardy, Italy (data from Italian government, Dipartimento di Protezione Civile available at \href{https://github.com/pcm-dpc/COVID-19}{https://github.com/pcm-dpc/COVID-19}); see Figure \ref{fig:growths-1}, left panel.\footnote{Since the number of infections is not observed, we match the growth rates of infection in the model with the growth rate of deaths in the data. This is justified when, as we assumed, the case fatality rate is constant and Death follows infection after a constant lag on average.} We chose to match data from Lombardy, the location where one of the first large clusters of infections occurred.
We assume that contagion rates are twice as large at Home locations than elsewhere and that Symptomatic individuals remain contagious at Home but are not contagious elsewhere.

We calibrate the daily fatality rates by type imposing that the Old have 10 times higher
fatality rates than the Young and scale their average so that the aggregate infection fatality rate (IFR) is 1 percent, consistently with, e.g., \cite{ferguson2020report}.

\subsubsection*{Transitions away and between the infected states, $A, Y, D, R$.}  

The probability to transition away from the Asymptomatic  state $A$, is $\rho + \nu$. We assumed  agents are infective only  in state A (we assumed that all sYmptomatic agents reduce to zero social contacts). The average time an agents stays in state  $A$ is then $T_{inf}=\frac{1}{\rho+\nu}$. We set  $\rho+\nu$ to match a theoretical moment which holds exactly at the initial condition in the basic SIR model. Recall $R_{0}$ denotes the number of agents a single infected agent at $t=0$ infects, on average. Let $g_0$ 
 denote the growth rate of the number of infected agents at $t=0$. Then, in SIR, 
    \setlength{\abovedisplayskip}{3pt}
      \setlength{\belowdisplayskip}{3pt}
\begin{equation}
\frac{(\mathcal{R}_{0}-1)}{T_{inf}}=g_0.\label{g}
\end{equation}
For the current SARS-CoV-2 epidemics, $R_{0}$ is reasonably estimated
between $2.5$ and $3.5$.\footnote{For details, see Footnote 32 in \cite{bisinmoro2020Geo}} The daily rate of growth of the infection $g$ is estimated $=.35$
by \cite{kaplan2020gianluca}.\footnote{\cite{alvarez2020simple} have .2; \cite{ferguson2020report}
have 0.15.} This implies, from Equation (\ref{g}), that $T_{inf}$ is between 4 and 7 days
(respectively for $\mathcal{R}_{0}$ between 2.5 and 3.5).  \cite{ferguson2020report} use 6.5 days. We set $\rho+\nu=.14$, so that $T_{inf}=\frac{1}{\rho+\nu}=7$.\footnote{We thank Gianluca Violante for suggesting this calibration strategy.}
Furthermore,
the average time from infection to death or recovery is reasonably
estimated to be 20 days; see e.g., \cite{ferguson2020report}. Therefore we set $\rho=.05$
so that $1 \rho=20$. This implies $\nu=.09$. 

The case-fatality
rate, the probability of death if infected, is estimated between .005
and .01; see e.g., \cite{ferguson2020report}.
Since agents remain sYmptomatic in the model, before Recovering, on
average $\frac{1}{\nu}=11$ days, we set the probability of Death
for a sYmptomatic, $\delta$, to be $0.001$.\footnote{We assume fatalities cannot occur to an agent less than 3 days before she becomes sYmptomatic and that every infected individual recovers with certainty after 100 days.}

\subsubsection*{Behavioral responses} We set $\phi=0.88$ as estimated by
\cite{keppo2020behavioral} using Swine flu data, and 
assume people start responding to the spread of the contagion very soon by setting $\underline{I}=0.01$. 
The fraction of firms that choose to let workers work remotely is calibrated  using the same functional form (eqn. \ref{Ketal}). We use the same calibration for $\phi=0.88$, 
but we calibrate $\underline{I}$ so that the fraction of School/Work locations that choose to shut down is 50 percent when 
the fraction of Asymptomatics is 20 percent (close to the peak of the Asymptomatics curve in most of our simulations\footnote{Solving
$(x/0.20)^{0.12}=0.5$ gives $x=0.00062$, therefore we use this value}). We believe
50 percent to be a reasonable guess for the peak given that 24 percent  of the population is reported by the 2018 ACS to be in school and that \citet{doepke20} compute from ATUS data that about 25 percent of workers work in highly telecommutable jobs; see also \citet{DingelNeiman20}, who report that the share of jobs that can be done at home goes from 50 percent to 30 percent, across metropolitan areas in the U.S.

\subsubsection*{Initial conditions} 
At time $t=0$ we set 30 individuals in Asymptomatic state; all 
others are Susceptibles. the Asymptomatics at $t= 0$ are those initially located in 
a position closest to City location $[x=0.25,y=0.25]$. We assume they have
been infected in the City. All others are Susceptibles. 


\newpage 
\bibliographystyle{aer}
\bibliography{covid}

\end{document}